%% file: sample-sigconf-authordraft.tex
\documentclass[sigconf]{acmart}

\usepackage{acronym}
\usepackage{subcaption}

\AtBeginDocument{%
  }
\DeclareGraphicsExtensions{.pdf,.png,.svg,.jpg,.mps}
    \input{acronyms}

\begin{document}

\title{EEG-Based Cognitive Load Classification During Landmark-Based VR Navigation}


\author{Jiahui An}
\orcid{0009-0008-0602-3904}
\affiliation{%
  \institution{Institute of Neuroinformatics, University of Zurich and ETH Zurich}
  \city{Zurich}
  \country{Switzerland}
}

\author{Bingjie Cheng}
\orcid{0009-0007-4930-1540}
\affiliation{%
  \institution{Department of Geography, University of Zurich}
  \city{Zurich}
  \country{Switzerland}}

\author{Dmitriy Rudyka}
\orcid{0009-0004-5450-6004}
\affiliation{%
  \institution{Department of Psychology, University of Zurich}
  \city{Zurich}
  \country{Switzerland}
}

\author{Elisa Donati}
\orcid{0000-0002-8091-1298}
\affiliation{%
  \institution{Institute of Neuroinformatics, University of Zurich and ETH Zurich}
  \city{Zurich}
  \country{Switzerland}}

\author{Sara Fabrikant}
\orcid{0000-0003-1263-8792}
\affiliation{%
  \institution{Department of Geography and Digital Society Initiative, University of Zurich}
  \city{Zurich}
  \country{Switzerland}}


\settopmatter{printacmref=false} 
\renewcommand\footnotetextcopyrightpermission[1]{} 
\pagestyle{plain} 
\renewcommand{\shortauthors}{} 
\thanks{Preprint version. This work is currently under review.}

\begin{abstract}
Brain–computer interfaces enable real-time monitoring of cognitive load, but their effectiveness in dynamic navigation contexts is not well established. Using an existing \ac{VR} navigation dataset, we examined whether \ac{EEG} signals can classify cognitive load during map-based wayfinding and whether classification accuracy depends more on task complexity or on individual traits. \ac{EEG} recordings from forty-six participants navigating routes with 3, 5, or 7 map landmarks were analyzed with a nested cross-validation framework across multiple machine learning models. Classification achieved mean accuracies up to 90.8\% for binary contrasts (3 vs.\ 7 landmarks) and 78.7\% for the three-class problem, both well above chance. Demographic and cognitive variables (age, gender, spatial ability, working memory) showed no significant influence. These findings demonstrate that task demands outweigh individual differences in shaping classification performance, highlighting the potential for task-adaptive navigation systems that dynamically adjust map complexity in response to real-time cognitive states.
\end{abstract}

\maketitle
\section{Introduction}
\label{sec:intro}
Spatial learning, which involves navigating through physical space, is a fundamental component of human cognition. It enables individuals to encode landmarks, form spatial representations, and remember directions.  
Effective spatial learning is essential in daily life, whether navigating urban centers, commuting through public transport systems, or exploring unfamiliar places while traveling. However, in today's digital society, people increasingly rely on GPS-based mobile maps to guide their navigation.
Although such systems offer efficiency and convenience, a growing body of research suggests that they negatively affect spatial learning performance~\cite{dahmani2020,mckinlay2016,parush2007, Ishikawa2019,Ruginski2019,sugimoto2021,munzer2012}. By providing turn-by-turn instructions, mobile maps reduce engagement with environmental cues and lead to weaker acquisition of survey knowledge and landmark memory~\cite{Ishikawa2019,Ruginski2019}. Understanding the cognitive processes involved in map-assisted navigation is therefore critical for designing adaptive human–machine systems that support both efficient wayfinding and spatial learning.

Navigation can be conceptualized as a dual-task paradigm: locomotion through the environment is cognitively straightforward, while the secondary task of constructing an internal representation of the environment is demanding~\cite{Montello2005}. 
Landmarks are central to this process, serving as salient environmental anchors that structure mental representations and support route memory and decision making~\cite{Richter2014}. However, processing landmarks requires cognitive resources, and excessive numbers can impose excessive mental load. Research indicates that while a moderate number of landmarks may aid learning, higher numbers can overwhelm users, increasing intrinsic cognitive load and impairing performance~\cite{Sweller1998, Cheng2023}. 

Empirical work confirms this trade-off~\cite{Cheng2022b,Cheng2023}, finding that spatial learning performance (landmark recognition and route memory) improved when the number of on-map landmarks increased from three to five, but did not improve further with seven landmarks. This behavioral evidence for an optimal landmark `sweet spot' was supported by \ac{EEG} data showing increased frontal theta \ac{ERS} and parieto-occipital P3 amplitude in the 7-landmark condition compared to the 5-landmark one, consistent with elevated cognitive load. These findings demonstrate that while landmarks are indispensable, excessive information can undermine spatial learning. Nevertheless, the analysis was limited to group-level comparisons of predefined biomarkers~\cite{Klimesch1999,do2021,delaux2021,wang2018,fu2006}. Thus, the relationship between moment-to-moment cognitive load states and spatial learning outcomes remains unquantified at the individual level, and the potential of a richer, multivariate neural signature to discriminate subtler states of cognitive overload remains unexplored. 

These observations align with cognitive load theory, which defines load as the mental effort required for information processing~\cite{Sweller1988}. In navigation, locomotion is relatively automatic, while wayfinding demands higher-level cognitive operations such as attention, working memory, and spatial reasoning. This dual demand increases cognitive load, particularly when users must attend to environmental cues while simultaneously consulting a mobile map. The design of navigation systems can either mitigate or exacerbate this load. Conventional GPS-based maps focus attention on the device, increasing extraneous load and reducing opportunities for spatial learning~\cite{Gardony2013,Ishikawa2019}. By contrast, including landmarks in digital maps can improve spatial knowledge acquisition, but excessive or poorly structured landmark information risks overloading the user~\cite{Cheng2023, Griffin2024}. Thus, striking the right balance is essential for designing navigation adaptive systems that support both efficient wayfinding and long-term spatial learning.

Traditional methods for assessing cognitive load in navigation have relied on dual-task paradigms and self-report instruments such as the NASA-TLX \cite{Hart1988}. However, dual-task methods can interfere with natural navigation, while self-reports offer only retrospective and subjective judgments that lack temporal resolution~\cite{Cheng2022a,Griffin2024}. These limitations highlight the need for direct, unobtrusive, and temporally precise measures of cognitive load in ecological navigation tasks. Building on these challenges, quantifying cognitive load during navigation in real-world settings is difficult due to uncontrollable factors such as traffic, weather, and individual behaviors (e.g., walking speed)~\cite{Darken2002, Gramann2013}. \ac{VR} offers a powerful methodological solution by combining ecological validity with experimental control. Immersive 3D environments allow researchers to replicate naturalistic navigation while systematically manipulating variables such as the number of landmarks on a map—conditions that are difficult to achieve in the real world~\cite{Sanchez-Vives2005, Klug2022}. Furthermore, \ac{VR} is fully compatible with neurophysiological recording methods such as \ac{EEG}, enabling the synchronous capture of brain activity and behavior during active navigation~\cite{Gramann2013, delaux2021}. Thus, a \ac{VR}-based paradigm provides an effective platform for investigating the cognitive load of landmark processing.

Recent work has investigated diverse physiological measures for cognitive load detection, such as \ac{HRV}, \ac{ET}, \ac{fNIRS}, \ac{EDA}, and \ac{EEG}~\cite{Wang2024, Qin2024, Aksu2024, An2025a, Anders2024}. Reliable real-time assessment of cognitive load is particularly important in \ac{HCI}, education, healthcare, and aviation~\cite{Dehais2020,Di2018,Kosch2023,Longo2018}. Among these modalities, \ac{EEG} is especially suitable because it captures brain activity at millisecond resolution, enabling the tracking rapid changes in mental effort without disrupting ongoing behavior.  Well-established neural signatures include increases in frontal theta and decreases in parietal alpha power~\cite{Klimesch1999,Gevins2003,do2021,delaux2021}. Building on these foundations, \ac{EEG} has been applied across domains such as education, driving, and aviation, demonstrating its value for real-time workload monitoring~\cite{Cheng2023,Di2018,Hassan2024,Anders2024,Wittke2025,An2025b,An2025c,Balakrishna2024}. In navigation, it further enables the tracking of cognitive demand fluctuations as users interact with mobile maps.

Advances in \ac{ML} have further expand \ac{EEG}’s potential. By combining multivariate features such as spectral band power, temporal dynamics, and statistical descriptors, classifiers including \ac{LR}, \ac{LDA}, \ac{SVM}, \ac{RF}, \ac{SNN}, \ac{XGBoost} and \ac{ANN} can distinguish between different cognitive states~\cite{Antonenko2010,Oh2014,Adeli2003a,Anders2024, Wittke2025, Nuamah2017, An2025b, An2025c, Ding2020}. These methods have achieved promising accuracy in controlled paradigms such as the n-back and go/no-go tasks, as well as in semi-naturalistic yet controlled settings such as driving or reading, demonstrating the feasibility of \ac{EEG}-based cognitive load classification. However, their application to ecologically valid navigation tasks remains limited.

A central pursuit in \ac{HCI} research is the design of adaptive systems that sense user states and adjust interfaces in real time to improve usability, preventing overload, and enhance task performance~\cite{Fischer2001,Jameson2007,Kosch2023}. Early approach followed a user-adaptive paradigm, tailoring interaction to relatively stable individual profiles such as demographics, expertise, user preferences and cognitive ability~\cite{Jameson2007,Fischer2001,Gevins2003,Darzi2021,Dehais2020,Longo2018}. More recent frameworks emphasize task- or state-adaptive approaches, which dynamically adjust interfaces in response to momentary task demands or user states~\cite{Dubiel2022,Yanez2025,Kosch2023}. Physiological adaptation systems exemplify this shift, modifying task difficulty based on affective or workload classification~\cite{Darzi2021}.

In navigation, it remains unclear whether \ac{EEG}-based classification is driven more by stable user traits or by the dynamic complexity of the task environment. Resolving this distinction is essential for designing adaptive systems that balance efficiency with spatial learning in realistic contexts.

This work builds on a \ac{VR} navigation paradigm developed for studying cognitive load~\cite{Cheng2022b}, from which the dataset was collected. Previous studies of digital map navigation analyzed EEG spectral power and \ac{ERP}s~\cite{Cheng2022b,Cheng2023}, but emphasized group-level effects. To our knowledge, no prior research has applied \ac{ML} to detect cognitive load from \ac{EEG} in digital map-based navigation. Here, we address this gap by evaluating classifiers across EEG channel subsets, moving beyond group-level averages toward trial-by-trial decoding of load induced by varying landmark quantities.

Our analysis addresses two research questions:
\begin{itemize}
    \item RQ1. Can EEG-based ML classifiers reliably distinguish levels of cognitive load during VR navigation?
    \item RQ2. Is variation in classifier performance better explained by individual traits (e.g., age, gender, working memory, spatial ability, perspective-taking) or by task complexity?
\end{itemize}

These comparisons directly inform the debate between user-adaptive and state-adaptive approaches in HCI. Methodologically, the study establishes a pipeline for EEG-based load classification in realistic navigation contexts. Conceptually, it provides empirical evidence on whether adaptive navigation systems should prioritize stable user profiles or dynamic detection of cognitive states.

\section{Methods}
\label{sec:methods}

\subsection{Dataset and Experimental Paradigm}
\label{ssec:dataset}
\paragraph{Participants:} Forty-seven participants (29 female, 18 male; age range = 18–35 years, M = 25.6, SD = 4.09) were recruited for a study on spatial learning and cognitive load during virtual navigation~\cite{Cheng2022b}. All provided written informed consent in accordance with ethical guidelines from the University's Ethics Board, the Swiss Psychological Society, and the American Psychological Association. Each participant received compensation of 30 CHF.
\\ Individuals with a history of neurological or psychiatric disorders were excluded. One participant (ID 51) was removed due to incomplete data, yielding a final sample of 46. Full details of the dataset and behavioral results are reported in the precursor study~\cite{Cheng2022b}.

\begin{figure}[t]
\centering
\includegraphics[width=\linewidth]{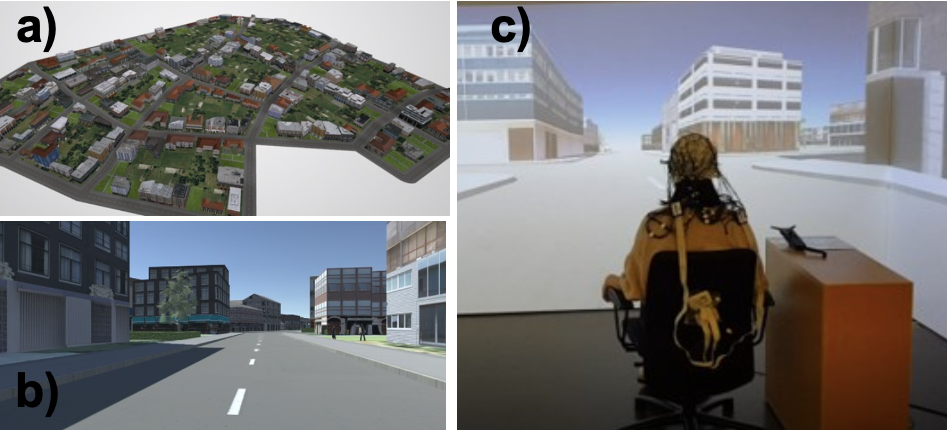}
\caption{Virtual navigation setup. Experimental setup. (a) Bird’s-eye view of a virtual city layout used for navigation. (b) First-person street-level perspective during navigation. (c) Participant seated in a three-wall \ac{CAVE} environment while wearing an \ac{EEG} cap. The \ac{CAVE} projects high-resolution 3D images onto three surrounding walls, with the display perspective continuously updated based on head position and orientation, providing an immersive navigation experience during \ac{EEG} recording.~\cite{Cheng2023}.}
\label{fig:vr_setup}
\Description{Panel a shows a top-down map of a virtual city with streets and buildings. 
Panel b shows a first-person street-level view with buildings and a tree along a road. 
Panel c shows a participant sitting in a CAVE room wearing an EEG cap, surrounded by projected city streets on three walls.}

\end{figure}

Participants navigated predefined routes in three distinct virtual European-style cities, designed in ArcGIS CityEngine 2018.0 and displayed in a stereoscopic~\ac{CAVE} system using Unity 2018.4 LTS (Fig.~\ref{fig:vr_setup}). Participants navigated from a first-person perspective while seated and instrumented for \ac{EEG}.

In this work, we analyzes \ac{EEG} data from a map-assisted navigation task. A within-participants design was used to examine how the number of landmarks (3, 5, or 7) displayed on a mobile map influenced cognitive load. These quantities were chosen visuospatial working memory research~\cite{Luck1997,Cowan2001,Baddeley2003,alvarez2004}, which estimates a core capacity of about four items, with higher limits for meaningful real-world objects such as landmarks~\cite{Baddeley2003, Brady2016, Brady2019}. This design allowed us to probe low (3), medium/high (5), and high/overload (7) cognitive load conditions. The precursor study confirmed the validity of this manipulation: the 7-landmark condition elicited significantly greater cognitive load, reflected in increased frontal theta power and P3 amplitude relative to the 3- and 5-landmark conditions.

\begin{figure}[t]
\centering
\includegraphics[width=\linewidth]{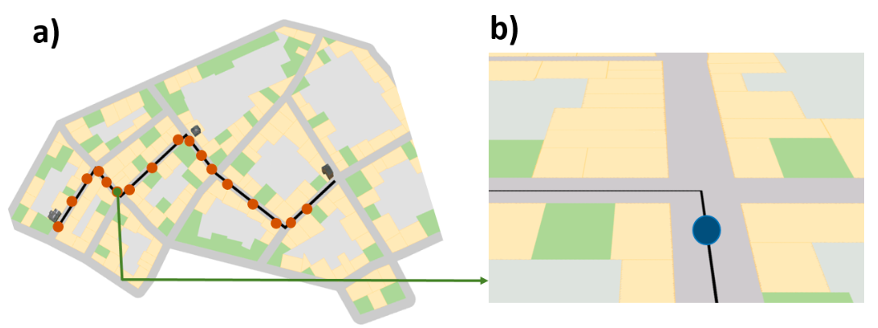}
\caption{Map pop-up procedure. a) 17 pop-up locations (red dots) along the predefined route. b) Example pop-up map indicating the participant’s current location (blue dot) and the upcoming route segment. The number of displayed landmark (3/5/7) varied according to assigned condition.}
\label{fig:map_popup}
\Description{Panel a) shows a route map with 17 red dots marking pop-up locations along a path through a city grid. 
Panel b) shows an example pop-up map with a blue dot marking the participant’s position and a black line indicating the next route segment, with nearby buildings shown.}

\end{figure}

\paragraph{Task and Procedure.}
Participants were instructed to reach a destination as quickly as possible using a rotating map displayed on the central screen. The map provided turn-by-turn instructions and indicated the participant's current location and heading. It appeared for 5-second intervals at 17 predefined points along each route (before and after intersections, and on straight segments) (Fig.~\ref{fig:map_popup}). During these intervals, the virtual environment was hidden and navigation was paused, simulating the real-world behavior of stopping to consult a mobile device. This design required participants to rely on memory for route continuation and landmark learning between map appearances.

\paragraph{Landmark Manipulation (Independent Variable).}

\begin{figure}[ht]
\centering
\includegraphics[width=8.5cm]{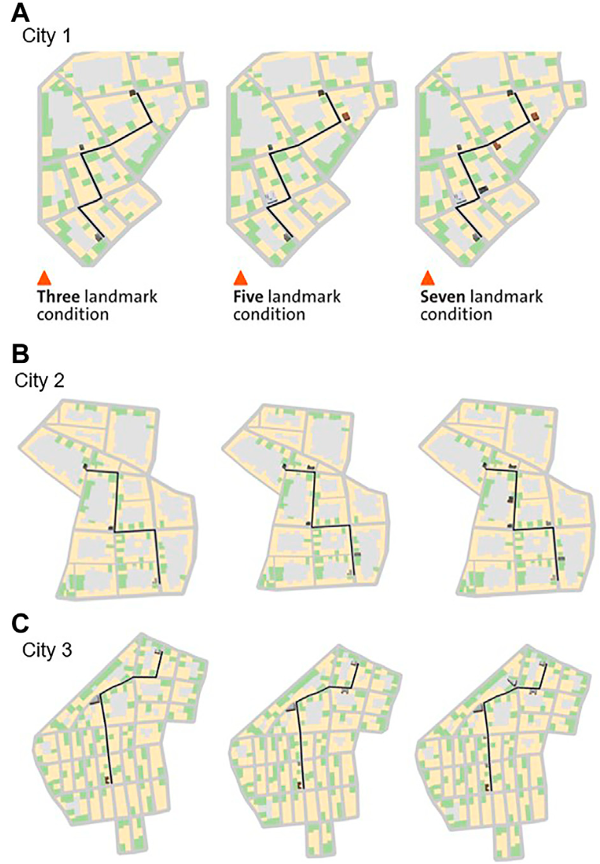}
\caption{Landmark-density conditions. Example mobile maps for the three conditions (3, 5, 7 landmarks). Conditions were counterbalanced across cities.}
\label{fig:landmark_conditions}
\Description{Panels with simplified mobile maps illustrating the three landmark densities (3, 5, 7).}
\end{figure}

The independent variable was the number of landmarks displayed on the map. Visually salient buildings at intersections were selected and rendered in 3D as landmarks, based on the criteria of persistence, salience, and informativeness~\cite{Stankiewicz2007}. Landmark positions were chosen to ensure equal spatial distribution along the route. Each participant experienced all three conditions, counterbalanced across cities to control order effects:

\begin{itemize}
\item 3-landmark condition
\item 5-landmark condition
\item 7-landmark condition
\end{itemize}
On each pop-up map, either 3, 5, or 7 landmarks were displayed, as shown in Fig.~\ref{fig:landmark_conditions}.

After navigating each city, participants' spatial knowledge was assessed through a battery of tests including landmark recognition, route direction, and a \ac{JRD} task. The analysis of these behavioral measures is reported in the precursor study~\cite{Cheng2023} and is not the focus of the current analysis. Responses were collected using a 3D responding and pointing device (WorldViz Inc, USA).

\paragraph{Cognitive Measures.}
Individual differences in spatial navigation ability and visuospatial cognitive capacity were assessed using established instruments:
\begin{itemize}
\item The \ac{SBSOD} \cite{Hegarty2002}, a self-report questionnaire measuring spatial orientation abilities and navigational style.
\item The \ac{PTSOT} \cite{Kozhevnikov2001}, which assesses the ability to imagine and orient from different spatial perspectives, with performance scored as angular error (in degrees).
\item The Corsi Block-Tapping Task \cite{Kessels2000}, a visual-spatial test of working memory capacity, scored as the maximum span length (partial score) a participant can correctly recall.
\end{itemize}

\subsection{EEG Acquisition and Preprocessing}
\ac{EEG} was recorded using a 64-channel LiveAmp system (Brain Products GmbH) with active electrodes arranged in an extended 10–20 montage. The signal was referenced to FCz, grounded at Fpz, and sampled at 500 Hz. Impedances were kept below 10 k$\Omega$. Data were streamed wirelessly via a UBT21 Bluetooth adapter and synchronized with navigation events (e.g., map onsets/offsets) using interprocess communication under Windows. 

\ac{EEG} data were preprocessed and analyzed using MNE-Python. Signals were bandpass filtered between 0.5–45\,Hz and notch filtered at 50\,Hz and harmonics to remove the line noise. Bad channels were identified and excluded, and data were re-referenced to the average reference. Excluded channels were reconstructed via interpolation. Epochs were extracted from –1\,s to +5\,s around each “showMap” event, with a baseline correction applied over –1\,s to 0\,s interval. \ac{ICA} was performed using the FastICA algorithm~\cite{Hyvarinen1999}, retaining components explaining 99\% of variance. Artifactual components (ocular, muscular, cardiac, line noise, channel noise) were automatically identified with ICLabel and subsequently removed.

\section{Feature Extraction}
\label{feature}

From the signal a total of 143 features were extracted. The decision to extract a broad set of features from temporal, spectral, and temporal–spectral domains was motivated by prior work~\cite{Wittke2025} showing that different signal characteristics capture complementary aspects of cognitive processing. Temporal features reflect variability and dynamics of ongoing neural activity, spectral features highlight oscillatory processes such as theta and alpha rhythms linked to attention and working memory, and temporal–spectral features capture transient changes that often occur during demanding cognitive states. As Wittke et al.~\cite{Wittke2025} argue, using a rich feature set maximizes the likelihood of identifying discriminative neural markers in ecologically valid tasks, where noise and inter-individual variability may otherwise obscure relevant patterns. To mitigate risks of overfitting in such high-dimensional spaces, we adopted \ac{ML} models with built-in regularization alongside cross-validation approach. We used \ac{ML} models that have been repeatedly shown to handle complex physiological data effectively ~\cite{Ding2020} and provide interpretable insights into feature contributions. In addition, we compared multiple model families, including linear classifiers and neural architectures to ensure that our results were not biased by the inductive biases of any single approach. The use of nested cross-validation further ensured that hyperparameter tuning was separated from performance estimation, which is a critical step for generalization in \ac{HCI} contexts where end-user deployment requires robustness across unseen individuals and environments.

\textbf{Temporal features (13 features):} Channel-averaged statistics derived from the raw \ac{EEG}, including amplitude measures (mean, \ac{SD}, minimum, maximum, peak), signal complexity metrics (outlier ratio, root mean square, zero-crossing rate), distribution descriptors (skewness, kurtosis), and Hjorth parameters (activity, mobility, complexity).

\textbf{Spectral features (22 features):} Band power was computed using Welch’s method (2-second Hanning windows, 50\% overlap) to capture five canonical frequency ranges: $\delta$ (0.5–4\,Hz), $\theta$ (4–8\,Hz), $\alpha$ (8–13\,Hz), $\beta$ (13–30\,Hz), and $\gamma$ (30–45\,Hz). Derived features included global inter-hemispheric asymmetry indices ($\delta$\_asy, $\theta$\_asy, $\alpha$\_asy, $\beta$\_asy, $\gamma$\_asy), regional asymmetries (e.g., frontal $\alpha$\_asy, frontotemporal indices), and within-channel ratios such as $\alpha/\theta$, $\theta/\alpha$, $\alpha/\beta$, $\theta/\beta$, $(\theta+\alpha)/\beta$, and $(\theta+\alpha)/(\alpha+\beta)$. Asymmetry was defined as:
\[
\mathrm{Asymmetry} = \ln(P_\text{right} + \epsilon) - \ln(P_\text{left} + \epsilon),
\]
with $\epsilon = 10^{-12}$ ensuring numerical stability. 

\textbf{Temporal–spectral features (108 features):} To capture the spatial distribution of the spectral power, the band power and ratios per channel were aggregated within each epoch. For each of nine spectral metrics ($\delta$, $\theta$, $\alpha$, $\beta$, $\gamma$, $\alpha/\beta$, $\theta/\beta$, $(\theta+\alpha)/\beta$, $(\theta+\alpha)/(\alpha+\beta)$), we computed 12 statistical descriptors across channels: mean, \ac{SD}, minimum, maximum, median, \ac{IQR}, coefficient of variation, skewness, kurtosis, Shannon entropy, absolute peak value and outlier ratio (fraction of channels exceeding 5\,SD).

The features were extracted using a sliding window approach (2 s duration with 50\% overlap) within each epoch, following methodologies established in recent literature~\cite{Wittke2025,Anders2024} while extending traditional feature sets with enhanced frequency domain and spatial aggregation metrics.

\subsection{Machine Learning Pipeline}
\subsubsection*{Classifiers and Hyperparameter Tuning}

We evaluated five supervised classifiers: \ac{LR}, \ac{SVM}, \ac{RF}, \ac{XGBoost}, and \ac{MLP}. Before training, all features were standardized using \texttt{StandardScaler}.  

For each model, the hyperparameters were optimized via a grid search within a nested cross-validation framework. The following parameter spaces were explored:  
\begin{itemize}
    \item \textbf{\ac{LR}}: Regularization strength (C), penalty type (l1, l2), solver.
    \item \textbf{\ac{SVM}}: Kernel (linear, rbf, poly), C, gamma, degree.
    \item \textbf{\ac{RF}}: Number of estimators (n\_estimators), maximum tree depth (max\_depth).
    \item \textbf{\ac{XGBoost}}: max\_depth, learning\_rate, subsample, colsample\_bytree, reg\_lambda.
    \item \textbf{\ac{MLP}}: Hidden layer sizes, activation function, L2 regularization (alpha).
\end{itemize}

\subsubsection*{Training and Evaluation Procedure}

A participant-wise nested stratified cross-validation was implemented to ensure robust generalizability and to prevent information leakage. For each participant, the available epochs were partitioned using an outer loop of 2 to 5 times (depending on the balance of the class). One fold served as the test set, while the remaining folds were reserved for training and hyperparameter optimization.  

Within each outer training set, an inner cross-validation loop (also 2 to 5 folds) was conducted to select the optimal hyperparameters from the model-specific grids. The best configuration was then retrained on the entire outer training set and evaluated in the corresponding outer test set. This procedure was repeated across all folds, ensuring unbiased estimation of model performance.  
We considered four classification tasks:  
\begin{enumerate}
    \item Multiclass: 3 vs.\ 5 vs.\ 7 landmarks,  
    \item Binary: 3 vs.\ 5 landmarks,  
    \item Binary: 3 vs.\ 7 landmarks,  
    \item Binary: 5 vs.\ 7 landmarks.  
\end{enumerate}

In summary, this nested cross-validation framework provided an unbiased evaluation of each classifier's ability to distinguish cognitive load levels under varying landmark conditions, while maintaining separation of training and test data within each participant. Unlike prior \ac{EEG} studies in navigation, which relied on aggregated statistical tests, our approach uses supervised \ac{ML} to decode cognitive load on a per-trial basis. 

Performance was assessed using multiple metrics: overall mean accuracy, macro-averaged F1-score, and per-class precision, recall, and F1-scores. Results were aggregated across outer folds to report mean and \ac{SD} for each metric. In addition, averaged maximum accuracy across outer folds (\texttt{cv\_max}) was recorded for each participant-model-task combination to capture potential upper-bound performance. 

The macro-averaged F1-score was employed because it provides a more robust evaluation than accuracy in the presence of potential class imbalance across cross-validation folds. Unlike accuracy, which weights all predictions equally, the Macro F1 calculates the F1-score (the harmonic mean of precision and recall) independently for each class and then takes the arithmetic mean. This ensures that each class contributes equally to the final metric, preventing majority classes from dominating the performance assessment and providing a more comprehensive measure of a model's ability to distinguish between all cognitive states.

Finally, to investigate the contribution of electrode coverage to classification performance, we repeated the full \ac{ML} pipeline using three different sets of \ac{EEG} channels:

\begin{itemize}
    \item All channels: full electrode montage after preprocessing.
    \item Frontal subset: electrodes over the frontal lobe, capturing activity associated with executive and working memory processes.
    \item Frontal–parietal subset: electrodes spanning both frontal and parietal cortices, targeting regions implicated in attentional control and working memory load.
\end{itemize}

To statistically assess differences between channel subsets, we compared paired participant-level scores using both paired t-tests and Wilcoxon signed-rank tests. Analyses were conducted separately for each task contrast (3-class classification, 3 vs. 5, 3 vs. 7, 5 vs. 7). Multiple testing was accounted for by reporting both p-values and effect sizes.

\subsection{Analysis of Individual Differences in Classification Performance}

To examine whether demographic and cognitive factors influenced which binary classification task yielded the best performance for each participant, we conducted a series of statistical analyses. Normality of continuous variables was assessed using the Shapiro–Wilk test.

In summary, we performed the following analyses:
\begin{enumerate}
    \item $\chi^2$ test of independence between gender and optimal task classification
    \item One-way ANOVAs for normally distributed variables across task groups (with gender as a covariate)
    \item Kruskal–Wallis tests for non-normally distributed variables
    \item Correlation analyses between classification accuracy and continuous variables (Pearson, gender-weighted, and partial correlations controlling for gender)
    \item Multinomial logistic regression predicting optimal task classification
\end{enumerate}




\subsubsection*{Group Comparison Tests}
We conducted parametric (ANOVA) or non-parametric (Kruskal-Wallis) tests based on data normality to examine differences in cognitive measures across participants grouped by their optimal classification task. The ANOVA model included task as the independent variable and cognitive scores as dependent variables, with gender included as a covariate.

\subsubsection*{Data Preparation and Weighting}
To address gender imbalance in the sample (28 female, 18 male participants), we calculated gender weights for each participant using:
\begin{equation}
w_g = \frac{N_{\text{total}}}{2 \times N_g}
\end{equation}
where $N_{\text{total}}$ represents the total sample size and $N_g$ represents the number of participants of gender $g$. These weights were applied in subsequent weighted analyses to ensure balanced representation.

\subsubsection*{Correlation Analyses}
We computed both regular Pearson correlations and gender-weighted correlations between classification accuracy and continuous variables of interest (age, \ac{SBSOD} spatial ability score, \ac{PTSOT} perspective-taking error, and Corsi working memory score). Weighted correlations were calculated using the DescrStatsW class from the statsmodels package which incorporates case weights in correlation computation.

For each continuous variable, we also calculated partial correlations with classification accuracy while controlling for gender effects. The partial correlation coefficient $r_{12.3}$ was computed using:
\begin{equation}
r_{12.3} = \frac{r_{12} - r_{13} \times r_{23}}{\sqrt{(1 - r_{13}^2)(1 - r_{23}^2)}}
\end{equation}
where $r_{12}$ represents the correlation between accuracy and the variable of interest, $r_{13}$ represents the correlation between accuracy and gender, and $r_{23}$ represents the correlation between the variable of interest and gender.

\subsubsection*{Multinomial Logistic Regression}

We implemented a multinomial logistic regression model to predict participants' optimal classification task:
\begin{equation}
\text{Task} \sim \text{Age} + \text{Gender} + \text{SBSOD} + \text{PTSOT error} + \text{Corsi score}
\end{equation}

Gender was encoded numerically (Male=1, Female=0), and task categories were encoded using label encoding. The model was fit using maximum likelihood estimation. Model performance was evaluated using classification accuracy, confusion matrices, and comparison against null accuracy.

All analyses were conducted using Python 3.9 with statsmodels, scikit-learn, and SciPy libraries. Statistical significance was evaluated at $\alpha = 0.05$, and all tests were two-tailed.

\section{Results}
\subsection{Model Performance Across Classification Tasks}

The nested cross-validation evaluation revealed distinct performance patterns across the four classification tasks and five machine learning models. Table ~\ref{tab:max_performance} summarizes the mean maximum cross validation accuracy (\texttt{cv\_max}) for each model-task combination, representing the upper bound of achievable performance across cross-validation folds.
\begin{table}[h]
\centering
\caption{Maximum Attainable Performance Across Classification Tasks (Mean $\pm$ SD, \%)}
\label{tab:max_performance}
\begin{tabular}{@{}llcc@{}}
\toprule
Task and Model & & \multicolumn{1}{l}{Accuracy} & \multicolumn{1}{l}{Macro F1} \\
& & \multicolumn{1}{l}{(\texttt{mean cv\_max}, \%)} & \\
\midrule
\multicolumn{4}{@{}l}{\textit{3-class classification}} \\
\quad XGBoost & & 78.7\% $\pm$ 13.7\% & 61.2\% $\pm$ 12.9\% \\
\quad RF & & 78.6\% $\pm$ 12.3\% & 61.2\% $\pm$ 12.8\% \\
\quad LR & & 77.9\% $\pm$ 12.6\% & 59.5\% $\pm$ 13.3\% \\
\quad MLP & & 75.8\% $\pm$ 13.1\% & 58.7\% $\pm$ 13.1\% \\
\quad SVM & & 74.7\% $\pm$ 14.2\% & 55.7\% $\pm$ 13.4\% \\
\addlinespace
\multicolumn{4}{@{}l}{\textit{3 vs. 5 classification}} \\
\quad XGBoost & & 90.8\% $\pm$ 10.4\% & 72.5\% $\pm$ 13.4\% \\
\quad MLP & & 90.4\% $\pm$ 10.0\% & 71.9\% $\pm$ 13.7\% \\
\quad RF & & 89.6\% $\pm$ 10.5\% & 72.5\% $\pm$ 13.9\% \\
\quad LR & & 88.7\% $\pm$ 10.8\% & 72.1\% $\pm$ 14.3\% \\
\quad SVM & & 87.8\% $\pm$ 13.6\% & 70.0\% $\pm$ 16.6\% \\
\addlinespace
\multicolumn{4}{@{}l}{\textit{3 vs. 7 classification}} \\
\quad RF & & 89.6\% $\pm$ 10.9\% & 73.6\% $\pm$ 14.1\% \\
\quad XGBoost & & 89.5\% $\pm$ 12.9\% & 72.0\% $\pm$ 13.5\% \\
\quad LR & & 89.0\% $\pm$ 12.1\% & 73.3\% $\pm$ 14.7\% \\
\quad SVM & & 88.8\% $\pm$ 12.5\% & 71.8\% $\pm$ 16.3\% \\
\quad MLP & & 87.8\% $\pm$ 11.4\% & 71.1\% $\pm$ 13.7\% \\
\addlinespace
\multicolumn{4}{@{}l}{\textit{5 vs. 7 classification}} \\
\quad LR & & 89.8\% $\pm$ 11.1\% & 71.5\% $\pm$ 13.6\% \\
\quad MLP & & 89.5\% $\pm$ 10.4\% & 71.5\% $\pm$ 11.1\% \\
\quad XGBoost & & 89.1\% $\pm$ 10.9\% & 71.4\% $\pm$ 12.4\% \\
\quad SVM & & 88.7\% $\pm$ 12.8\% & 68.6\% $\pm$ 15.8\% \\
\quad RF & & 88.5\% $\pm$ 12.2\% & 72.9\% $\pm$ 12.2\% \\
\bottomrule
\end{tabular}
\end{table}

\begin{figure}[ht]
\centering
\begin{subfigure}{0.48\linewidth}
    \centering
    \includegraphics[width=\linewidth]{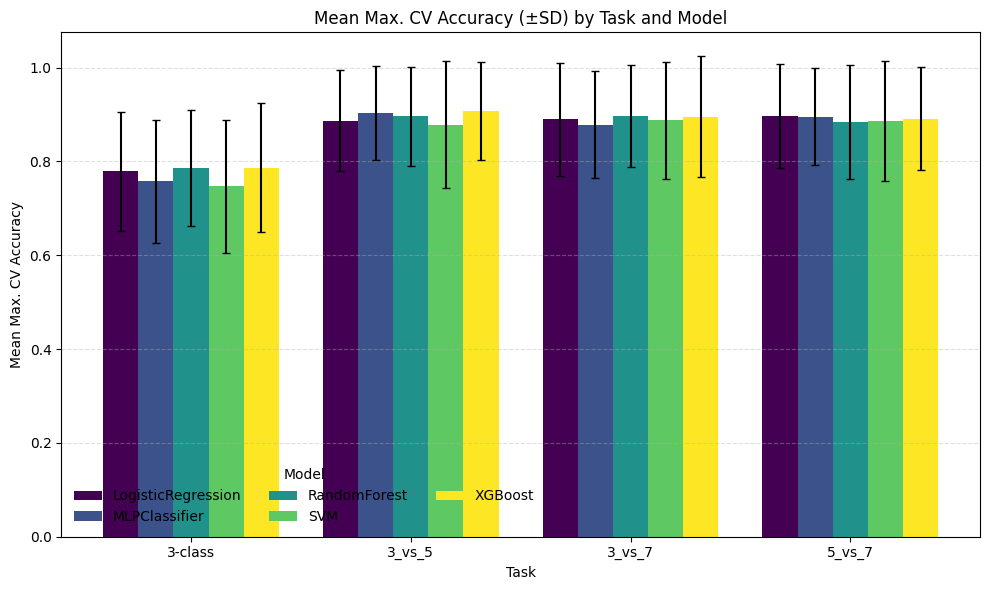}
    \caption{Mean maximum CV accuracy across tasks and models.}
    \label{fig:performance_accuracy}
\end{subfigure}
\hfill
\begin{subfigure}{0.48\linewidth}
    \centering
    \includegraphics[width=\linewidth]{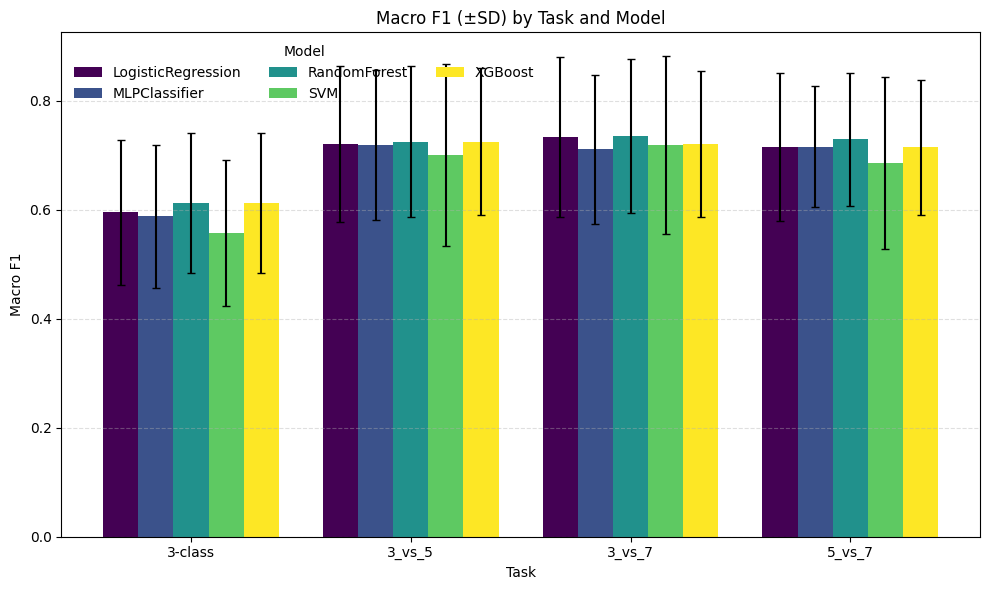}
    \caption{Macro F1 scores across tasks and models.}
    \label{fig:performance_f1}
\end{subfigure}
\caption{Comparison of classification performance across tasks and models. (a) Mean maximum cross validation accuracy across participants, (b) macro F1 scores. Error bars represent standard deviations.}
\label{fig:performance_comparison}
\Description{Two grouped bar charts. (a) Mean maximum cross-validation accuracy across four models: LR, MLP, RF, SVM, and XGBoost compared across tasks: 3-class, 3 vs 5, 3 vs 7, and 5 vs 7. Binary tasks achieve higher accuracy than the 3-class condition, with all models performing similarly within each task. (b) Macro F1 scores across the same models and tasks show the same pattern: binary tasks outperform the 3-class condition, with small differences between models. Error bars indicate standard deviations.}

\end{figure}

The nested cross-validation results showed a consistent performance hierarchy across models (Fig.~\ref{fig:performance_comparison}). Binary classification tasks achieved higher accuracy than the multiclass condition, with the 3 vs.\ 7 landmarks task yielding the highest performance. Maximum cross-validation accuracy reached 90.8\% for the best model–task combination (Fig.~\ref{fig:performance_accuracy}). Macro F1 scores exhibited the same pattern (Fig.~\ref{fig:performance_f1}).

\subsubsection*{Key Findings}
The \texttt{cv\_max} metric, representing the maximum accuracy across outer folds, indicated that individual participants could achieve higher performance (up to 89-91\% accuracy) than the average cross-validated results, suggesting significant potential for optimal model configuration though participant-specific factors may still influence optimal model performance. Moreover, the \ac{SD} across folds were typically between 0.10 to 0.15, indicating considerable variability in performance across different data partitions.

Classification was strongest for binary tasks, peaking at 90.8\% for 3 vs. 5 landmarks, but performance remained robust in the more complex 3-class problem (78.7\%). Thus, while finer gradations of load are more challenging to distinguish, they remain feasible. Across models, tree-based approaches (\ac{RF} and \ac{XGBoost}) performed best, followed by \ac{LR}, \ac{MLP}, and \ac{SVM}. 

To further examine whether such performance patterns related to individual traits, we assessed distributions of participant measures. Shapiro–Wilk tests indicated that all variables except \ac{PTSOT} error were normally distributed across task groups (p > 0.05), whereas \ac{PTSOT} error significantly deviated from normality (p < 0.05). Analysis of the 46 participants further revealed a comparable distribution of optimal binary classification types: 18 participants (39.1\%) performed best on the 3 vs. 5 landmarks task, 14 (30.4\%) on the 3 vs. 7 task, and 14 (30.4\%) on the 5 vs. 7 task.


\subsubsection*{Analysis of Demographic and Cognitive Variables}

One-way ANOVAs controlling for gender showed no significant differences in age ($p = 0.769$), spatial ability ($p = 0.485$), or working memory capacity ($p = 0.465$) across participants grouped by their optimal classification task. Non-parametric Kruskal–Wallis test revealed no significant differences in perspective-taking error ($p = 0.954$). 

A marginal association was observed between gender and optimal task, $\chi^2(2, N=46)=5.948$, $p=0.051$. Female participants more often showed optimal performance on the 3 vs.\ 5 task (50.0\%), whereas male participants more often favored the 3 vs.\ 7 task (50.0\%).

Descriptive statistics are reported in Table~\ref{tab:demographic_vars} and inferential results are reported in Table ~\ref{tab:individual_differences}.

\begin{table}[h]
\centering
\caption{Descriptive statistics (mean ± SD) for demographic and cognitive variables by optimal classification task}
\label{tab:demographic_vars}
\begin{tabular}{@{}lcccc@{}}
\toprule
Variable & \multicolumn{1}{l}{3 vs. 5} & \multicolumn{1}{l}{3 vs. 7} & \multicolumn{1}{l}{5 vs. 7} & \multicolumn{1}{l}{$p$} \\
& \multicolumn{1}{l}{(n=18)} & \multicolumn{1}{l}{(n=14)} & \multicolumn{1}{l}{(n=14)} & \\
\midrule
Age & 24.94 $\pm$ 4.72 & 25.36 $\pm$ 3.27 & 26.00 $\pm$ 3.90 & 0.769 \\
SBSOD & 4.91 $\pm$ 0.96 & 4.46 $\pm$ 1.31 & 4.76 $\pm$ 0.85 & 0.485 \\
PTSOT & 26.33 $\pm$ 21.20 & 27.18 $\pm$ 23.84 & 21.29 $\pm$ 11.22 & 0.954 \\
Corsi & 4.33 $\pm$ 0.61 & 4.12 $\pm$ 0.50 & 4.13 $\pm$ 0.55 & 0.465 \\
\bottomrule
\end{tabular}
\end{table}

\begin{table}[h]
\centering
\caption{Inferential test results summary for demographic and cognitive variables across task groups}
\label{tab:individual_differences}
\begin{tabular}{lc}
\toprule
Analysis & Result \\
\midrule
Gender–Task Association & $\chi^2(2, N=46) = 5.948, p = 0.051$ \\
Age by Task & $F(2,43) = 0.264, p = 0.769$ \\
SBSOD by Task & $F(2,43) = 0.736, p = 0.485$ \\
PTSOT Error by Task & $\chi^2(2, N=46) = 0.095, p = 0.954$ \\
Corsi Score by Task & $F(2,43) = 0.780, p = 0.465$ \\
\bottomrule
\end{tabular}
\end{table}

\subsubsection*{Correlational and Predictive Analyses}

Classification accuracy showed no significant correlations with age ($r=0.124$, $p=0.411$), spatial ability ($r=-0.048$, $p=0.752$), perspective-taking error ($r=-0.171$, $p=0.255$), or working memory ($r=0.003$, $p=0.984$). These null results remained consistent when applying gender weights, conducting gender-stratified analyses, and controlling for gender via partial correlations (all $|r| < .20$).


A multinomial logistic regression model predicting optimal task classification achieved limited accuracy (47.8\%), only marginally exceeding the null accuracy rate (39.1\%). No demographic or cognitive variables emerged as significant predictors in the model.

\begin{figure*}[ht]
    \centering
    \includegraphics[width=\textwidth]{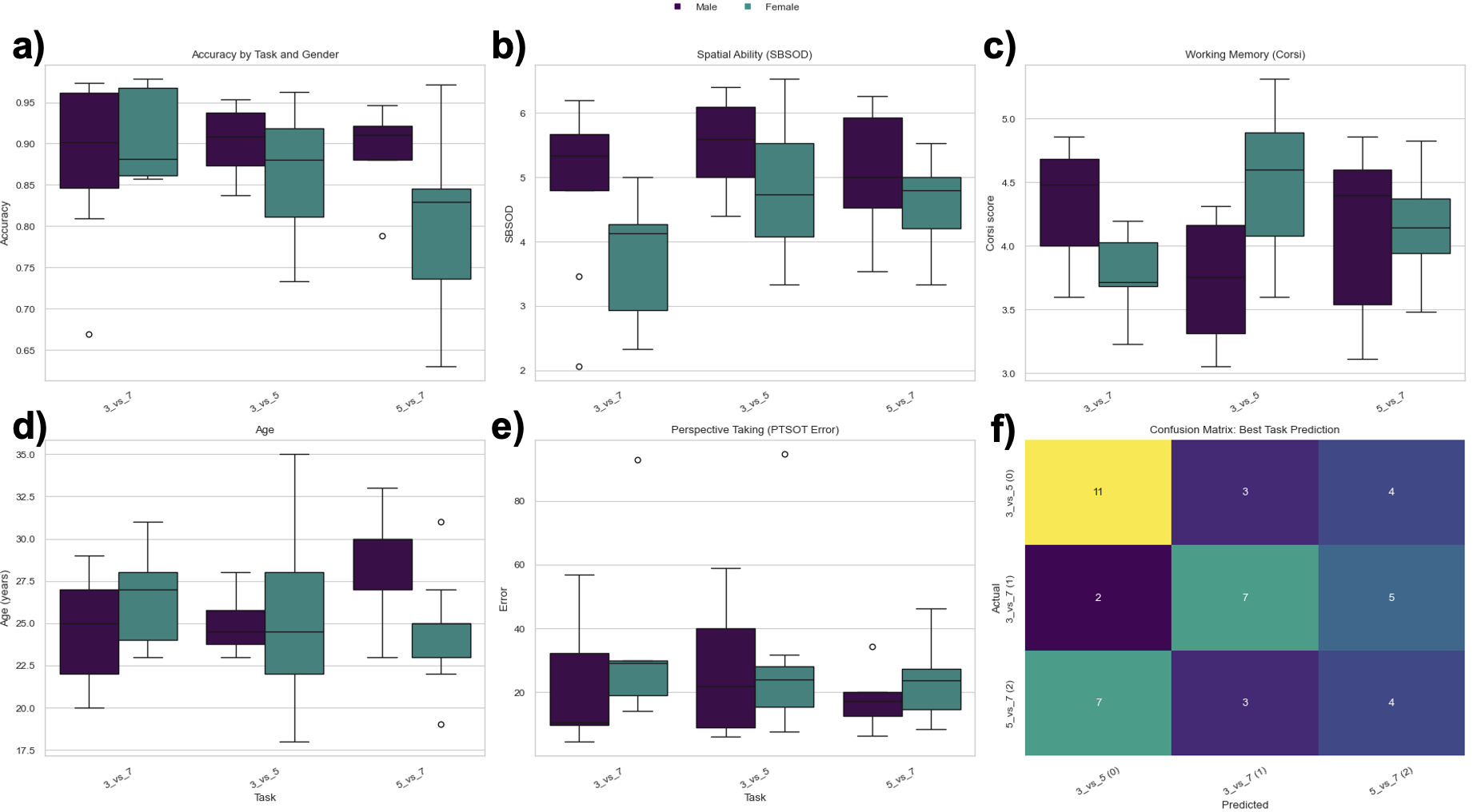}
    \caption{Visualization of demographic and cognitive variables across task groups and genders, and confusion matrix of multinomial logistic regression. Boxplots show distributions of (a) classification accuracy, (b) spatial ability \ac{SBSOD}, (c) working memory (Corsi score), (d) age, and (e) perspective-taking error \ac{PTSOT} across gender and optimal classification task. These plots corroborate the statistical findings: group differences are small and largely overlapping, and model predictions only marginally exceeded the null baseline (f) Confusion matrix illustrates predictive performance of the multinomial logistic regression model.}
    \label{fig:demog_gender}
\end{figure*}

Our analyses revealed a consistent pattern: demographic and cognitive variables showed no significant predictive power for optimal classification performance. A marginal trend with gender was observed ($p = 0.051$), but overall, classifier performance was shaped primarily by task demands rather than stable participant traits.

Fig.~\ref{fig:demog_gender} illustrates these relationship. Panels a)–e) show gender-stratified boxplots for accuracy, spatial ability, working memory, age, and perspective-taking error across task groups. Although inferential tests did not reveal significant group differences, several weak, non-significant trends are visible: a) men showed slightly higher median accuracies in the 3 vs.\ 7 and 5 vs.\ 7 tasks; b) spatial ability (\ac{SBSOD}) indicated a small male advantage in the 3 vs.\ 7 group; c) working memory fluctuated inconsistently across tasks and genders; d) the 5 vs.\ 7 group skewed marginally older; and e) males tended to have lower \ac{PTSOT} error in the 3 vs.\ 7 and 5 vs.\ 7 tasks, though with high variability. Panel f) shows the confusion matrix of the multinomial logistic regression, which reached only modest accuracy ($\approx$48\%), slightly above the null baseline, with most correct classifications in the 3 vs.\ 5 group and frequent confusions between 5 vs.\ 7 and 3 vs.\ 5.


\begin{figure}[ht]
\centering
\includegraphics[width=\linewidth]{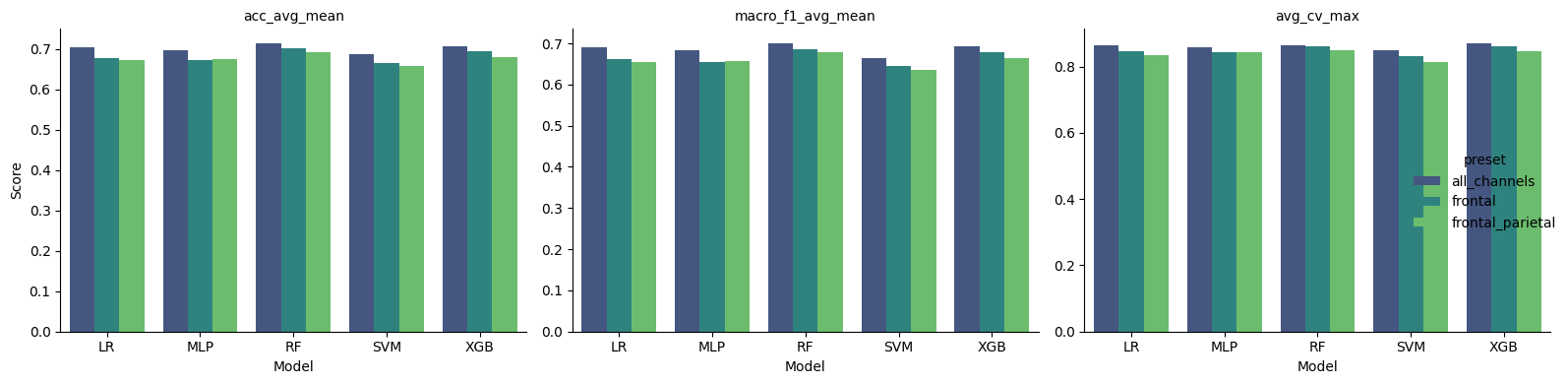}
\caption{Model performance across electrode subsets. Bars show mean accuracy, macro-F1, and cv\_max for each classifier using all channels, frontal, and frontal–parietal presets. This figure highlights the relative performance hierarchy of channel sets across metrics.}
\label{fig:subset_overview}
\Description{Three grouped bar plots comparing performance across models (LR, MLP, RF, SVM, XGBoost). Each plot shows mean accuracy, macro-F1, and cv\_max for three electrode presets: all channels, frontal, and frontal–parietal. Bars for all channels are consistently highest, frontal slightly lower, and frontal–parietal lowest across most metrics.}

\end{figure}

\begin{figure}[ht]
\centering
\includegraphics[width=\linewidth]{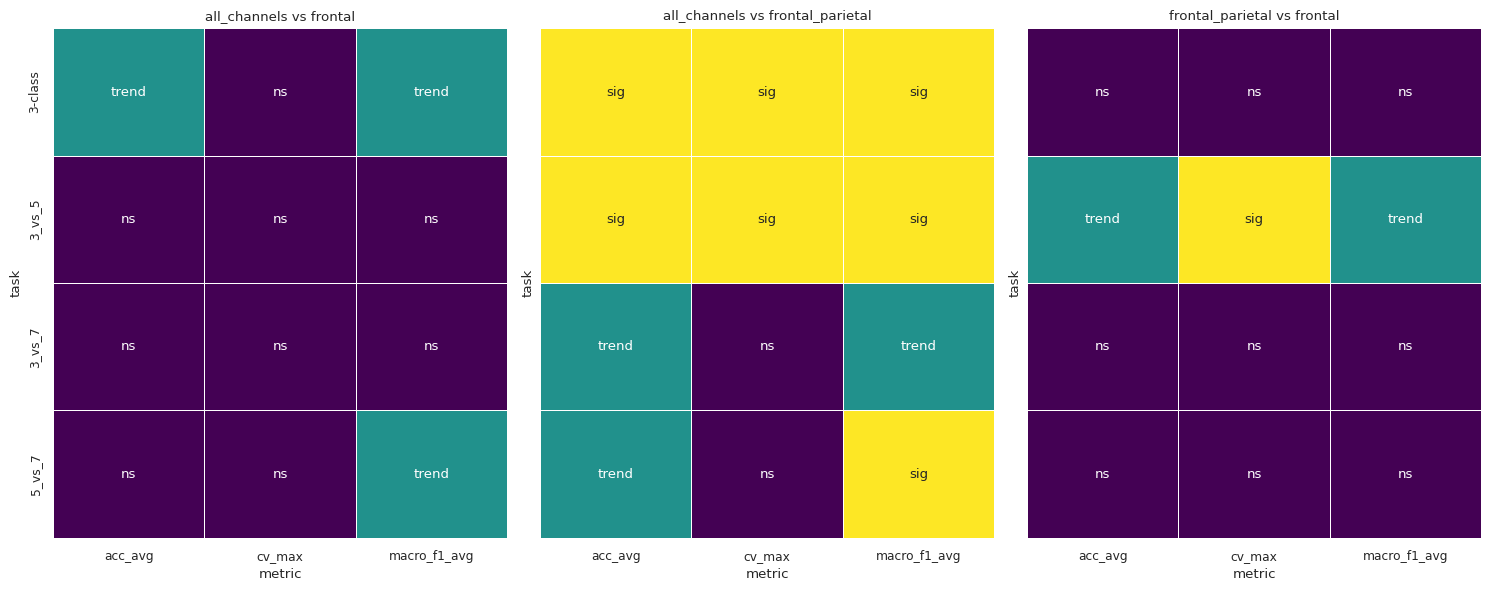}
\caption{Statistical comparison of presets by task. Each panel summarizes pairwise tests across tasks and metrics: \emph{all vs.\ frontal} (left), \emph{all vs.\ frontal–parietal} (middle), and \emph{frontal–parietal vs.\ frontal} (right). Cells encode outcomes (ns, trend, sig); the all-channels montage significantly outperforms frontal–parietal in most 3-class/3vs5 metrics, while frontal often remains comparable to all in the binary tasks.}
\label{fig:subset_stats}
\Description{Three heatmaps comparing electrode subsets across tasks and metrics. Each heatmap panel summarizes pairwise tests: all channels vs frontal (left), all channels vs frontal–parietal (middle), frontal–parietal vs frontal (right). Cells are labeled ns, trend, or sig. Results show that all channels significantly outperform frontal–parietal in most 3-class and 3 vs 5 metrics, while frontal often performs comparably to all channels in binary tasks.}
\end{figure}

\subsection{The Impact of Electrode Coverage on Classification Performance}
We compared models trained on all 64 channels, a frontal-only subset, and a frontal–parietal subset. Across metrics and classifiers, the full montage consistently yielded the highest performance (Fig.~\ref{fig:subset_overview}). Restricting input to the frontal subset produced only minor accuracy losses ($\Delta \approx$ 1–2\%), while the frontal–parietal subset showed larger decrements ($\Delta \approx$ 2–3\%), particularly in the three-class condition where most differences reached statistical significance (p $<$ 0.01; Fig.~\ref{fig:subset_stats}). For binary contrasts (3 vs.\ 7, 5 vs.\ 7), the frontal subset achieved comparable results to the all-channel baseline (p $>$ 0.7), whereas the frontal–parietal configuration again underperformed. Interestingly, extending coverage to parietal sites did not provide systematic benefits and in some cases even reduced performance relative to frontal-only input.

These effects were stable across classifiers, with \ac{RF} and \ac{XGBoost} generally achieving the highest absolute scores, but following the same ranking of channel sets.

\section{Discussion}
\label{sec:discussions}

Addressing RQ1, our results demonstrate that \ac{EEG}-based classifiers can reliably distinguish levels of cognitive load, with accuracies reaching 78.7\% for three-class and up to 90.8\% for binary contrasts, underscoring the feasibility of EEG-based workload detection in ecologically valid but controlled \ac{VR} tasks. These values are competitive with, and in some cases exceed, performance reported in other domains. For example, simulated driving studies report similar accuracy ranges: Wang et al.~\cite{wang2023} achieved $\sim$90\% for two-class classification using deep neural networks, and Di Flumeri et al.~\cite{Di2018} reported $\sim$0.75 AUC for three-class classification when combining \ac{EEG} and \ac{ET}. Laboratory paradigms such as the n-back task typically yield accuracies in the 70–80\% range for three-class classification task (e.g., 77.2\% in~\cite{khanam2022}; 74.9\% in~\cite{nguyen2024}). By contrast, more unconstrained scenarios show marked reductions: Wittke et al.~\cite{Wittke2025} reported only $\sim$60\% F1 for digital textbook reading using a similar feature-extraction and ML pipeline. Together, these comparisons indicate that structured yet dynamic environments, such as VR navigation or driving simulators, offer the balance of ecological validity and experimental control needed to elicit robust neural signatures of cognitive load.

Our findings addressing RQ2 provide clear evidence that \ac{EEG}-based cognitive load classification during navigation is primarily task-driven rather than user-driven. Classifiers decoded load across landmark conditions with accuracies well above chance (Table~\ref{tab:max_performance}), yet performance was systematically determined by task complexity rather than individual characteristics. Neither demographic (age, gender) nor cognitive factors (working memory, spatial ability, perspective-taking) predicted classification outcomes (Table~\ref{tab:individual_differences}). This suggests that the dominant source of variance in neural signatures of cognitive load arises from the task environment itself rather than from stable user profiles.

Cognitive load during map-assisted VR navigation increases with the number of displayed landmarks: learning improves from 3 to 5 landmarks without additional load, but 7 landmarks elicit higher frontal theta and parietal P3 responses, reflecting cognitive overload without learning benefits~\cite{Cheng2022b}. Our findings extend this literature by showing that these neural responses are not only present but also discriminable at the level of individual trials. This provides direct support for the ``inefficient overload'' state described in earlier work, where adding landmarks increases neural effort without producing behavioral gains. Critically, classification accuracy exceeded 83\% even for the subtle 5 vs.\ 7 landmark contrast, a condition that is behaviorally indistinguishable. This demonstrates that overload is reflected in distributed, multivariate neural patterns spanning spectral, temporal, and time–frequency domains, rather than in isolated biomarkers. Whereas traditional univariate analyses captured only the most prominent components (e.g., frontal theta, parietal P3), our feature-rich approach revealed finer-grained, spatially distributed signatures that more comprehensively characterize overload states.

Unlike prior workload studies in education or driving, our work directly addresses digital navigation interfaces, an important \ac{HCI} application where adaptive interaction design can immediately benefit from reliable cognitive load detection. From this perspective, these results move the field from descriptive group-level analyses toward predictive, trial-level modeling of cognitive state. By showing that \ac{EEG} can track load dynamically in real time, our work lays a foundation for closed-loop neuro-adaptive systems. Such systems could, for example, regulate the number or salience of landmarks displayed on a digital map to prevent overload while maintaining wayfinding efficiency. Importantly, classifier performance was largely comparable across different algorithms (LR, MLP, RF, SVM, XGBoost) once task was controlled and with a rich set of features, indicating that task demands outweighed model choice in shaping performance. This finding suggests that designing adaptive systems should prioritize optimizing task conditions rather than seeking marginal gains from algorithmic tuning.

Electrode subset analyses further support the feasibility of lightweight, applied EEG systems. Frontal electrodes alone were sufficient to capture the neural signatures of cognitive load, particularly in binary contrasts, while adding parietal sites unexpectedly reduced performance. This may reflect additional noise or inter-individual variability introduced by parietal signals not directly aligned with load-sensitive activity. Although the full 64-channel montage produced the highest overall accuracies, the frontal-only configuration yielded statistically comparable performance for critical contrasts (e.g., high vs.\ overload). This is encouraging for applied neuroergonomics, as it demonstrates that compact, wearable EEG systems could achieve practical utility in real-world navigation contexts.

Several limitations should be acknowledged. First, the participant pool consisted of young, highly educated adults, limiting generalizability to older or more diverse populations~\cite{Hartshorne2015,vanDerHam2020}. Future work should examine whether similar cognitive capacity limits emerge in older adults, who may experience reduced working memory and slower attentional shifts, or in populations with different educational and cultural backgrounds. Second, the VR environment was constrained to medium-length, grid-like routes with track-up map orientation; results may differ in more organic city layouts, under north-up map orientations, or when participants navigate longer and more complex journeys. Third, we examined load during initial encoding in a novel environment, whereas load dynamics likely evolve with repeated exposure, consolidation, and retrieval.

Future research should address these limitations by testing more heterogeneous participant groups and more varied navigation environments. Another promising direction is multimodal integration. Prior studies show that combining \ac{EEG} with complementary measures such as \ac{ET}, \ac{EDA}, or \ac{HRV} substantially improves classification~\cite{Qin2024,Anders2024,Pusica2024,An2025c,Liu2025}. Recording gaze behavior alongside EEG could, for example, link neural signatures of load with fixation patterns, saccades, and pupil responses, producing more comprehensive and reliable estimates. Experimental manipulations could also amplify cognitive demand by varying route complexity, landmark salience, or time pressure, or by targeting retrieval phases such as spatial memory recall and intersection decision-making. Ultimately, advancing toward closed-loop paradigms will be essential: systems that detect overload in real time and dynamically adapt map design (e.g., simplifying routes, modulating landmark density) will allow direct evaluation of whether adaptive interventions improve both usability and spatial learning.

\section{Conclusion}
\label{sec:conclusion}
In conclusion, this study demonstrates that EEG-based cognitive load classification during navigation is primarily task-driven rather than user-driven. To our knowledge, it provides the first evidence that EEG signals collected during digital map navigation can be reliably classified into load states, with accuracies up to 90.8\% for optimal contrasts. Crucially, robust decoding was achievable even with reduced frontal montages, underscoring the feasibility of lightweight, portable systems for real-world use. These findings indicate that adaptive navigation aids should prioritize dynamic adjustment to task demands rather than static user profiling. Such real-time adaptation offers a pathway to preventing overload and enhancing spatial learning.

\section*{Acknowledgments}


We thank Sara Irina Fabrikant and Bingjie Cheng for providing access to the dataset and the opportunity to analyze it. This work was supported by the European Research Council (ERC) [Advanced Grant GeoViSense, No. 740426], the Swiss National Science Foundation (SNSF) [GeoNavLearn: Towards geographic context-adaptive mobile geographic information displays to support navigators’ spatial learning, No. 10003761], and the University of Zurich Digital Society Initiative (DSI) Excellence Program Fellowship.


\bibliographystyle{unsrt}
\bibliography{ref}

\appendix

\end{document}

%% file: acronyms.tex
\acrodef{AOI}[AOI]{Areas of Interest}
\acrodef{ATC}[ATC]{Air Traffic Control} 
\acrodef{ADC}[ADC]{Analog-to-Digital Converter}
\acrodef{ADEXP}[AdExp-IF]{Adaptive Exponential Integrate-and-Fire}
\acrodef{ADM}[ADM]{Asynchronous Delta Modulator}
\acrodef{AE}[AE]{Address-Event}
\acrodef{AER}[AER]{Address-Event Representation}
\acrodef{AEX}[AEX]{AER EXtension board}
\acrodef{AFE}[AFE]{Analog Front-End}
\acrodef{AFM}[AFM]{Atomic Force Microscope}
\acrodef{AGC}[AGC]{Automatic Gain Control}
\acrodef{AI}[AI]{Artificial Intelligence}
\acrodef{AMDA}[AMDA]{AER Motherboard with D/A converters}
\acrodef{AMPA}[AMPA]{$\alpha$-Amino-3-hydroxy-5-methyl-4-isoxazolepropionic Acid}
\acrodef{ANN}[ANN]{Artificial Neural Network}
\acrodef{API}[API]{Application Programming Interface}
\acrodef{APMOM}[APMOM]{Alternate Polarity Metal On Metal}
\acrodef{ARM}[ARM]{Advanced RISC Machine}
\acrodef{ASIC}[ASIC]{Application Specific Integrated Circuit}
\acrodef{BCM}[BMC]{Bienenstock-Cooper-Munro}
\acrodef{BD}[BD]{Bundled Data}
\acrodef{BEOL}[BEOL]{Back-end of Line}
\acrodef{BG}[BG]{Bias Generator}
\acrodef{BMI}[BMI]{Brain-Machince Interface}
\acrodef{BTB}[BTB]{Band-to-Band tunnelling}
\acrodef{bpm}[bpm]{Beats per Minute}
\acrodef{CA}[CA]{Cortical Automaton}
\acrodef{CAD}[CAD]{Computer Aided Design}
\acrodef{CAM}[CAM]{Content Addressable Memory}
\acrodef{CAVIAR}[CAVIAR]{Convolution AER Vision Architecture for Real-Time}
\acrodef{CCN}[CCN]{Cooperative and Competitive Network}
\acrodef{CDR}[CDR]{Clock-Data Recovery}
\acrodef{CFC}[CFC]{Current to Frequency Converter}
\acrodef{CHP}[CHP]{Communicating Hardware Processes}
\acrodef{CMIM}[CMIM]{Metal-Insulator-Metal Capacitor}
\acrodef{CML}[CML]{Current Mode Logic}
\acrodef{CMOL}[CMOL]{Hybrid CMOS nanoelectronic circuits}
\acrodef{CMOS}[CMOS]{Complementary Metal-Oxide-Semiconductor}
\acrodef{CNN}[CNN]{Convolutional Neural Network}
\acrodef{CNS}[CNS]{central Nervous System}
\acrodef{COTS}[COTS]{Commercial Off-The-Shelf}
\acrodef{CPG}[CPG]{Central Pattern Generator}
\acrodef{CPLD}[CPLD]{Complex Programmable Logic Device}
\acrodef{CPU}[CPU]{Central Processing Unit}
\acrodef{CSM}[CSM]{Cortical State Machine}
\acrodef{CSP}[CSP]{Constraint Satisfaction Problem}
\acrodef{CTXCTL}[CTXCTL]{CortexControl}
\acrodef{CV}[CV]{Coefficient of Variation}
\acrodef{DAC}[DAC]{Digital to Analog Converter}
\acrodef{DAS}[DAS]{Dynamic Auditory Sensor}
\acrodef{DAVIS}[DAVIS]{Dynamic and Active Pixel Vision Sensor}
\acrodef{DBN}[DBN]{Deep Belief Network}
\acrodef{DBS}[DBS]{Deep Brain Stimulation}
\acrodef{DFA}[DFA]{Deterministic Finite Automaton}
\acrodef{DIBL}[DIBL]{Drain-Induced Barrier-Lowering}
\acrodef{DI}[DI]{Delay Insensitive}
\acrodef{divmod3}[DIVMOD3]{Divisibility of a number by three}
\acrodef{DMA}[DMA]{Direct Memory Access}
\acrodef{DNF}[DNF]{Dynamic Neural Field}
\acrodef{DNN}[DNN]{Deep Neural Network}
\acrodef{DOF}[DOF]{Degrees of Freedom}
\acrodef{DPE}[DPE]{Dynamic Parameter Estimation}
\acrodef{DPI}[DPI]{Differential Pair Integrator}
\acrodef{DRAM}[DRAM]{Dynamic Random Access Memory}
\acrodef{DR}[DR]{Dual Rail}
\acrodef{DRRZ}[DR-RZ]{Dual-Rail Return-to-Zero}
\acrodef{DSP}[DSP]{Digital Signal Processor}
\acrodef{DVS}[DVS]{Dynamic Vision Sensor}
\acrodef{DYNAP-SE}[DYNAP-SE]{Dynamic Neuromorphic Asynchronous Processor}
\acrodef{EBL}[EBL]{Electron Beam Lithography}
\acrodef{ECG}[ECG]{Electrocardiography}
\acrodef{ECoG}[ECoG]{Electrocorticography}
\acrodef{EDA}[EDA]{Electrodermal activity}
\acrodef{EDVAC}[EDVAC]{Electronic Discrete Variable Automatic Computer}
\acrodef{EEG}[EEG]{Electroencephalography}
\acrodef{EI}[EI]{Excitatory-Inhibitory}
\acrodef{EIN}[EIN]{Excitatory-Inhibitory Network}
\acrodef{EM}[EM]{Expectation Maximization}
\acrodef{EMG}[EMG]{Electromyography}
\acrodef{EOG}[EOG]{Electrooculogram}
\acrodef{EPSC}[EPSC]{Excitatory Post-Synaptic Current}
\acrodef{EPSP}[EPSP]{Excitatory Post-Synaptic Potential}
\acrodef{EZ}[EZ]{Epileptogenic Zone}
\acrodef{FAA}[FAA]{frontal alpha asymmetry}
\acrodef{FDSOI}[FDSOI]{Fully-Depleted Silicon on Insulator}
\acrodef{FET}[FET]{Field-Effect Transistor}
\acrodef{FFT}[FFT]{Fast Fourier Transform}
\acrodef{FI}[F-I]{Frequency--Current}
\acrodef{FMA}[FMA]{Floating Microelectrode Array}
\acrodef{FNN}[FNN]{Feed-forward Neural Network}
\acrodef{FPGA}[FPGA]{Field Programmable Gate Array}
\acrodef{FR}[FR]{Fast Ripple}
\acrodef{FSA}[FSA]{Finite State Automaton}
\acrodef{FSM}[FSM]{Finite State Machine}
\acrodef{GABA}[GABA]{$\gamma$-Aminobutanoic Acid}
\acrodef{GIDL}[GIDL]{Gate-Induced Drain Leakage}
\acrodef{GOPS}[GOPS]{Giga-Operations per Second}
\acrodef{GPIO}[GPIO]{General Purpose I/O}
\acrodef{GPU}[GPU]{Graphical Processing Unit}
\acrodef{GSE}[GSE]{Gaze Stationary Entropy} 
\acrodef{GT}[GT]{Ground Truth}
\acrodef{GTE}[GTE]{Gaze Transition Entropy} 
\acrodef{GUI}[GUI]{Graphical User Interface}
\acrodef{HAL}[HAL]{Hardware Abstraction Layer}
\acrodef{HFO}[HFO]{High Frequency Oscillation}
\acrodef{HH}[H\&H]{Hodgkin \& Huxley}
\acrodef{HMM}[HMM]{Hidden Markov Model}
\acrodef{HR}[HR]{Heart Rate}
\acrodef{HRS}[HRS]{High-Resistive State}
\acrodef{HSD}[HSD]{Honest Significant Difference} 
\acrodef{HSE}[HSE]{Handshaking Expansion}
\acrodef{HW}[HW]{Hardware}
\acrodef{hWTA}[hWTA]{Hard Winner-Take-All}
\acrodef{HRV}[HRV]{hearth rate variability}
\acrodef{IC}[IC]{Integrated Circuit}
\acrodef{ICA}[ICA]{Indipendent Component Analysis}
\acrodef{ICT}[ICT]{Information and Communication Technology}
\acrodef{iEEG}[iEEG]{Intracranial Electroencephalography}
\acrodef{IF2DWTA}[IF2DWTA]{Integrate \& Fire 2-Dimensional WTA}
\acrodef{IF}[I\&F]{Integrate-and-Fire}
\acrodef{IFSLWTA}[IFSLWTA]{Integrate \& Fire Stop Learning WTA}
\acrodef{IMU}[IMU]{Inertial Measurement Unit}
\acrodef{INCF}[INCF]{International Neuroinformatics Coordinating Facility}
\acrodef{INI}[INI]{Institute of Neuroinformatics}
\acrodef{IO}[I/O]{Input/Output}
\acrodef{IoT}[IoT]{Internet of Things}
\acrodef{IP}[IP]{Intellectual Property}
\acrodef{IPSC}[IPSC]{Inhibitory Post-Synaptic Current}
\acrodef{IPSP}[IPSP]{Inhibitory Post-Synaptic Potential}
\acrodef{ISI}[ISI]{Inter-Spike Interval}
\acrodef{JFLAP}[JFLAP]{Java - Formal Languages and Automata Package}
\acrodef{LEDR}[LEDR]{Level-Encoded Dual-Rail}
\acrodef{LFP}[LFP]{Local Field Potential}
\acrodef{LIFE}[LIFE]{Longitudinal Intrafascicular Electrodes}
\acrodef{LIF}[LIF]{Leaky Integrate-and-Fire}
\acrodef{LLC}[LLC]{Low Leakage Cell}
\acrodef{LMS}[LMS]{Least Mean Squares}
\acrodef{LNA}[LNA]{Low-Noise Amplifier}
\acrodef{LPF}[LPF]{Low Pass Filter}
\acrodef{LR}[LR]{Logistic Regression}
\acrodef{LRS}[LRS]{Low-Resistive State}
\acrodef{LSM}[LSM]{Liquid State Machine}
\acrodef{LTD}[LTD]{Long Term Depression}
\acrodef{LTI}[LTI]{Linear Time-Invariant}
\acrodef{LTP}[LTP]{Long Term Potentiation}
\acrodef{LTU}[LTU]{Linear Threshold Unit}
\acrodef{LUT}[LUT]{Look-Up Table}
\acrodef{LVDS}[LVDS]{Low Voltage Differential Signaling}
\acrodef{MCMC}[MCMC]{Markov-Chain Monte Carlo}
\acrodef{MAE}[MAE]{Mean Absolute Error}
\acrodef{MEA}[MEA]{Multielectrode Arrays}
\acrodef{MEMS}[MEMS]{Micro Electro Mechanical System}
\acrodef{MFR}[MFR]{Mean Firing Rate}
\acrodef{MIM}[MIM]{Metal Insulator Metal}
\acrodef{ML}[ML]{Machine Learning}
\acrodef{MLP}[MLP]{Multilayer Perceptron}
\acrodef{monoNSM}[monoNSM]{Monotonic Neural State Machine}
\acrodef{MOSCAP}[MOSCAP]{Metal Oxide Semiconductor Capacitor}
\acrodef{MOSFET}[MOSFET]{Metal Oxide Semiconductor Field-Effect Transistor}
\acrodef{MOS}[MOS]{Metal Oxide Semiconductor}
\acrodef{MRI}[MRI]{Magnetic Resonance Imaging}
\acrodef{NCS}[NCS]{Neuromorphic Cognitive Systems}
\acrodef{NDFSM}[NDFSM]{Non-deterministic Finite State Machine}
\acrodef{ND}[ND]{Noise-Driven}
\acrodef{NEF}[NEF]{Neural Engineering Framework}
\acrodef{NHML}[NHML]{Neuromorphic Hardware Mark-up Language}
\acrodef{NIL}[NIL]{Nano-Imprint Lithography}
\acrodef{NI}[NI]{Neural Interface}
\acrodef{NMDA}[NMDA]{\textit{N}-Methyl-\textsc{d}-aspartate}
\acrodef{NME}[NE]{Neuromorphic Engineering}
\acrodef{NN}[NN]{Neural Network}
\acrodef{nnNSM}[nnNSM]{Nearest Neighbors Neural State Machine}
\acrodef{NOC}[NoC]{Network-on-Chip}
\acrodef{NRZ}[NRZ]{Non-Return-to-Zero}
\acrodef{NSM}[NSM]{Neural State Machine}
\acrodef{OR}[OR]{Operating Room}
\acrodef{OTA}[OTA]{Operational Transconductance Amplifier}
\acrodef{PCB}[PCB]{Printed Circuit Board}
\acrodef{PCHB}[PCHB]{Pre-Charge Half-Buffer}
\acrodef{PCM}[PCM]{Phase Change Memory}
\acrodef{PC}[PC]{Personal Computer}
\acrodef{PDK}[PDK]{Process Design Kit}
\acrodef{PE}[PE]{Phase Encoding}
\acrodef{PFA}[PFA]{Probabilistic Finite Automaton}
\acrodef{PFC}[PFC]{Prefrontal Cortex}
\acrodef{PFM}[PFM]{Pulse Frequency Modulation}
\acrodef{PGA}[PGA]{Programmable Gain Amplifier}
\acrodef{PNI}[PNI]{Peripheral Nerve Interface}
\acrodef{PNS}[PNS]{Peripheral Nervous System}
\acrodef{PPG}[PPG]{Photoplethysmography}
\acrodef{PR}[PR]{Production Rule}
\acrodef{PSC}[PSC]{Post-Synaptic Current}
\acrodef{PSD}[PSD]{Power spectral density}
\acrodef{PSP}[PSP]{Post-Synaptic Potential}
\acrodef{PSTH}[PSTH]{Peri-Stimulus Time Histogram}
\acrodef{PV}[PV]{Parvalbumin}
\acrodef{QDI}[QDI]{Quasi Delay Insensitive}
\acrodef{RAM}[RAM]{Random Access Memory}
\acrodef{RA}[RA]{Resected Area}
\acrodef{RDF}[RDF]{Random Dopant Fluctuation}
\acrodef{RELU}[ReLu]{Rectified Linear Unit}
\acrodef{RLS}[RLS]{Recursive Least-Squares}
\acrodef{RMSE}[RMSE]{Root Mean Square-Error}
\acrodef{RRMSE}[RRMSE]{Relative Root Mean Square Error}
\acrodef{RMS}[RMS]{Root Mean Square}
\acrodef{RNN}[RNN]{Recurrent Neural Network}
\acrodef{ROLLS}[ROLLS]{Reconfigurable On-Line Learning Spiking}
\acrodef{RRAM}[R-RAM]{Resistive Random Access Memory}
\acrodef{R}[R]{Ripple}
\acrodef{RBF}[RBF]{Radial basis function}
\acrodef{RISC}[RISC]{Reduced Instruction Set Computer}
\acrodef{RSA}[RSA]{Respiratory Sinus Arrhythmia}
\acrodef{SAC}[SAC]{Selective Attention Chip}
\acrodef{SAT}[SAT]{Boolean Satisfiability Problem}
\acrodef{SCI}[SCI]{Spinal Cord Injury}
\acrodef{SCX}[SCX]{Silicon CorteX}
\acrodef{SD}[SD]{Signal-Driven}
\acrodef{SEM}[SEM]{Spike-based Expectation Maximization}
\acrodef{SCR}[SCR]{Skin Conductance Response}
\acrodef{SLAM}[SLAM]{Simultaneous Localization and Mapping}
\acrodef{SNN}[SNN]{Spiking Neural Network}
\acrodef{SNR}[SNR]{Signal to Noise Ratio}
\acrodef{SOC}[SoC]{System-On-Chip}
\acrodef{SOI}[SOI]{Silicon on Insulator}
\acrodef{SOZ}[SOZ]{Seizure Onset Zone}
\acrodef{SP}[SP]{Separation Property}
\acrodef{SPI}[SPI]{Serial Peripheral Interface}
\acrodef{SRAM}[SRAM]{Static Random Access Memory}
\acrodef{SST}[SST]{Somatostatin}
\acrodef{STDP}[STDP]{Spike-Timing Dependent Plasticity}
\acrodef{STD}[STD]{Short-Term Depression}
\acrodef{STP}[STP]{Short-Term Plasticity}
\acrodef{STT-MRAM}[STT-MRAM]{Spin-Transfer Torque Magnetic Random Access Memory}
\acrodef{STT}[STT]{Spin-Transfer Torque}
\acrodef{SVM}[SVM]{Support Vector Machine}
\acrodef{SW}[SW]{Software}
\acrodef{sWTA}[sWTA]{soft Winner-Take-All}
\acrodef{TEMP}[TEMP]{Temperature}
\acrodef{TCAM}[TCAM]{Ternary Content-Addressable Memory}
\acrodef{TFT}[TFT]{Thin Film Transistor}
\acrodef{TIME}[TIME]{Transverse Intrafascicular Multichannel Electrode}
\acrodef{TLE}[TLE]{Temporal Lobe Epilepsy}
\acrodef{UEA}[UEA]{Utah Electrode Array}
\acrodef{USB}[USB]{Universal Serial Bus}
\acrodef{USEA}[USEA]{Utah Slanted Electrode Array}
\acrodef{VHDL}[VHDL]{VHSIC Hardware Description Language}
\acrodef{VHSIC}[VHSIC]{Very High Speed Integrated Circuits}
\acrodef{VIP}[VIP]{Vasoactive Intestinal Peptide}
\acrodef{VLSI}[VLSI]{Very Large Scale Integration}
\acrodef{VNS}[VNS]{Vagal Nerve Stimulation}
\acrodef{VOR}[VOR]{Vestibulo-Ocular Reflex}
\acrodef{VSA}[VSA]{Vector Symbolic Architecture}
\acrodef{WCST}[WCST]{Wisconsin Card Sorting Test}
\acrodef{WTA}[WTA]{Winner-Take-All}
\acrodef{XML}[XML]{eXtensible Mark-up Language}
\acrodef{ERS}[ERS]{Event-Related Synchronization}
\acrodef{VR}[VR]{Virtual Reality}
\acrodef{fNIRS}[fNIRS]{functional Near-Infra Red Spectroscopy}
\acrodef{HCI}[HCI]{Human-Computer Interaction}
\acrodef{LDA}[LDA]{Linear Discriminant Analysis}
\acrodef{RF}[RF]{Random Forests}
\acrodef{JRD}[JRD]{Judgement of Relative Direction}
\acrodef{PTSOT}[PTSOT]{Perspective Taking/Spatial Orientation Test}
\acrodef{SBSOD}[SBSOD]{Santa Barbara Sense of Direction Scale}
\acrodef{IQR}[IQR]{Interquartile Range}
\acrodef{SD}[SD]{Standard Deviation}
\acrodef{XGBoost}[XGBoost]{Extreme Gradient Boosting}
\acrodef{BFGS}[BFGS]{Broyden–Fletcher–Goldfarb–Shanno}
\acrodef{ET}[ET]{Eye-Tracking}
\acrodef{CAVE}[CAVE]{Cave Automatic Virtual Environment}
\acrodef{ERP}[ERP]{Event-Related Potential}